\documentclass[10pt,conference]{IEEEtran}
\IEEEoverridecommandlockouts

\usepackage[normalem]{ulem}
\usepackage{listings}
\usepackage{pgfplots, pgfplotstable}
\usepackage{color}
\usepackage{wrapfig}
\usepackage{booktabs}
\usepackage{url,hyperref,lineno,microtype,multirow}
\usepackage{graphicx}
\graphicspath{{figs/}}

\newcommand{\NAMENS} {\textsc{RBF}} 
\newcommand{\NAME} {\textsc{RBF }} 

\newcommand{\ignore}[1]{}

\usepackage{cite}
\usepackage{amsmath,amssymb,amsfonts}
\usepackage{algorithmic}
\usepackage{graphicx}
\usepackage{textcomp}
\usepackage{xcolor}
\def\BibTeX{{\rm B\kern-.05em{\sc i\kern-.025em b}\kern-.08em
    T\kern-.1667em\lower.7ex\hbox{E}\kern-.125emX}}

\begin{document}
\title{Hybrid Edge-HPC Systems for Low-Latency Data-Driven Inference}
\author{
    \IEEEauthorblockN{
	Liubov Kurafeeva\IEEEauthorrefmark{1},
        Ryan Hartung\IEEEauthorrefmark{2},
        Benjamin Carter\IEEEauthorrefmark{1},
        Alan Subedi\IEEEauthorrefmark{3},
        Avhishek Biswas\IEEEauthorrefmark{3},
        Michael Fay\IEEEauthorrefmark{3},\\
        Shantenu Jha\IEEEauthorrefmark{4}\IEEEauthorrefmark{5},
        Chandra Krintz\IEEEauthorrefmark{2}
        Andre Merzky\IEEEauthorrefmark{5},
        Douglas Thain\IEEEauthorrefmark{2},
        Memet Can Vuran\IEEEauthorrefmark{3},
        Rich Wolski\IEEEauthorrefmark{2}
    }
    \IEEEauthorblockA{
    \IEEEauthorrefmark{1}Dept. of Computer Science,\\
University of California Santa Barbara}
    \IEEEauthorblockA{
    \IEEEauthorrefmark{2}Dept. of Computer Science and Engineering\\
University of Notre Dame}
    \IEEEauthorblockA{
    \IEEEauthorrefmark{3}School of Computing\\
University of Nebraska-Lincoln}
    \IEEEauthorblockA{
    \IEEEauthorrefmark{4}Princeton Plasma Physics Laboratory}
    \IEEEauthorblockA{
    \IEEEauthorrefmark{5}Dept. of Electrical and Computer Engineering\\
Rutgers-New Brunswick}
}
\maketitle
\begin{abstract}
Emerging cyber-physical systems increasingly require low-latency inference
from streaming sensor data while maintaining models that reflect complex and
evolving physical processes. However, in many domains, model updates depend on
high-fidelity simulations and training that execute on remote high-performance
computing (HPC) systems under batch scheduling. This creates a fundamental
mismatch between the responsiveness required at the edge and the cost,
throughput, and availability of simulation-driven model updates.

To resolve this mismatch, we present \NAME (Reverse Backfill), 
a hybrid edge–HPC learning and inference
architecture that integrates low-latency edge inference with asynchronous,
simulation-driven model improvement. RBF targets simulation-bounded settings
in which model updates are limited by simulation throughput and HPC scheduling
delays, and reinterprets HPC backfilling by using opportunistic computation to
improve model accuracy rather than system utilization. \NAME decouples inference
from simulation and training by deploying lightweight surrogate models at the
edge while incorporating improved models asynchronously as they become
available. The architecture supports pluggable surrogate models and
orchestrates computation across heterogeneous infrastructure spanning edge
devices, private 5G networks, cloud resources, and HPC systems.

We instantiate \NAME using a real-world digital agriculture deployment that
couples edge sensing with computational fluid dynamics (CFD) simulations to
infer spatial airflow patterns in a large screenhouse. Our evaluation
characterizes end-to-end system behavior under realistic constraints,
quantifying simulation latency, training cost, inference throughput, and the
impact of delayed model updates on prediction accuracy. Results demonstrate
that RBF enables continuous, low-latency inference while improving model
fidelity over time, effectively integrating HPC resources into operational
cyber-physical systems despite delayed and irregular model updates.

\end{abstract}


\section{Introduction}
\label{sec:intro}

Emerging cyber-physical systems increasingly require low-latency
inference from streaming sensor data while maintaining models that
capture complex and evolving physical processes. In many domains,
including environmental monitoring, energy systems, and digital
agriculture, these models depend on high-fidelity simulations that
are computationally expensive and must execute on remote
high-performance computing (HPC) systems.

This creates a fundamental mismatch: edge applications require
continuous, low-latency inference, while model updates are governed
by simulation throughput and HPC scheduling delays. As a result,
systems must operate with stale models and improve them
opportunistically as new simulation results become available.

To enable this, we present \NAME (Reverse Backfill), a hybrid 
edge-HPC learning and
inference architecture. \NAME treats
HPC resources not as synchronous backends, but as asynchronous
model-improvement engines. Edge nodes perform continuous inference
using surrogate models, while simulation and training pipelines
generate improved models opportunistically as resources become
available. This design enables uninterrupted inference while
reducing model staleness over time.

\NAME targets simulation-driven settings in which model updates are
bounded by the cost and scheduling of high-fidelity computation.
This distinguishes it from data-driven continuous learning systems,
where models can be updated directly from streaming observations.
In \NAMENS, model improvement is inherently delayed and irregular,
requiring explicit system support for asynchronous updates.

The architecture is built around three key ideas. First, it decouples
inference from simulation-driven model updates, allowing edge nodes
to operate continuously while models are retrained asynchronously.
Second, it supports pluggable surrogate models and simulation
pipelines, enabling flexible tradeoffs among accuracy, latency, and
compute cost. Third, it orchestrates computation across the
edge-cloud-HPC continuum, integrating sensing, simulation, and
inference across heterogeneous infrastructure.

We instantiate \NAME in a real-world digital agriculture deployment
that couples edge sensing with computational fluid dynamics (CFD)
simulations to infer spatial airflow patterns in a large agricultural
screenhouse. This instantiation serves as a testbed for evaluating
the system’s behavior under realistic constraints, including HPC
queue delays and variable model update intervals.

Our evaluation focuses on system-level tradeoffs in
simulation-driven inference pipelines. We characterize the latency
and throughput of the simulation-training pipeline, evaluate the
impact of model staleness on prediction accuracy, and measure the
overhead of model transfer and edge inference. The results show that
system performance is dominated by simulation latency, and that
networking and inference overheads are negligible in comparison.

\begin{figure*}[t]
\includegraphics[width=\textwidth]{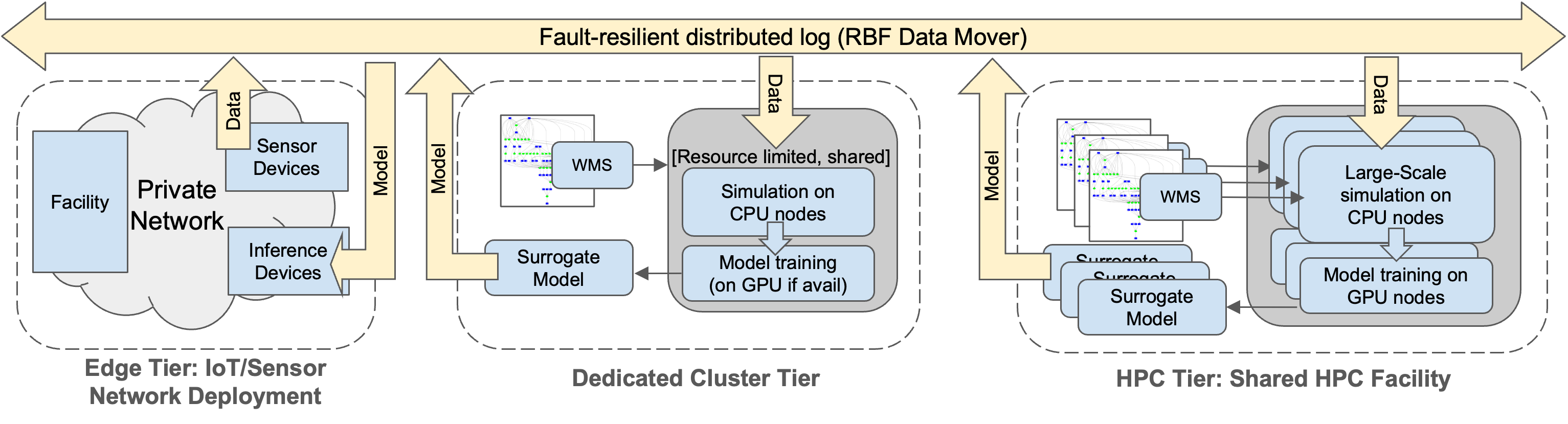}
\caption{\NAME System Architecture \label{fig:arch}}
{The Reverse Backfill architecture has three tiers.  
(Left) At the remote facility, sensor data is collected and
conveyed over a private wireless network to a fault resilient distributed log using a pub/sub protocol.  When
available, published models are
downloaded and used for rapid inference
in place of simulation.
(Middle) A dedicated, but resource-limited, cluster
pulls the published data, runs a simulation of the facility, and
 uses the output to train a surrogate model, which is published back to the distributed
log.  (Right) At a shared HPC facility, Reverse Backfilling is used to execute
 more simulations concurrently, albeit at a  less predictable cadence.  This augments the capacity of the
 dedicated cluster by filling in intermediate timepoints.
  }
\end{figure*}

In summary, this paper makes the following contributions:
\begin{itemize}
\item \textbf{Asynchronous HPC-in-the-loop operation.}
\NAME decouples low-latency inference from simulation-driven model
updates, enabling continuous inference under simulation-bounded
updates and incorporating improved models opportunistically.

\item \textbf{Pluggable hybrid modeling.}
\NAME supports interchangeable simulation and surrogate modeling
components, enabling flexible evaluation across model types and
system constraints.

\item \textbf{Distributed orchestration across the computing continuum.}
\NAME coordinates sensing, simulation, and inference across edge,
cloud, and HPC resources while maintaining responsiveness at the edge through 5G network slicing.
\end{itemize}
Taken together, this paper demonstrates how HPC resources can be integrated into
operational cyber-physical systems as asynchronous components in continuous
inference loops. Doing so enables on-demand decision support while preserving the
fidelity advantages of high-resolution simulation.

The remainder of the paper is organized as follows. 
Section~\ref{sec:arch} presents the
\NAME architecture, including its design goals, the learning and inference
loop, the surrogate training pipeline, and mechanisms for model orchestration
and deployment across edge and HPC resources. Section~\ref{sec:impl} describes an
instantiation of this architecture in a real-world agricultural screenhouse
deployment and digital agriculture application. 
Section~\ref{sec:eval} uses this deployment to empirically evaluate the
tradeoffs enabled by \NAME and to characterize its end-to-end performance
across the simulation, training, and inference pipeline.

\section{\NAME Architecture}
\label{sec:arch}

\NAME enables continuous inference in systems where model updates
are governed by high-fidelity simulation and delayed computation.
The architecture is designed to operate under four constraints:
(i) low-latency inference requirements at the edge,
(ii) high computational cost of simulation and training,
(iii) batch scheduling and queue delays on HPC systems, and
(iv) evolving environments that require periodic model updates.

\NAME is composed of three logical tiers, as illustrated 
in Figure~\ref{fig:arch}.
The edge tier performs
sensor data generation, ingestion, and low-latency inference. The
dedicated cluster tier executes simulation and training workloads
at a regular cadence. The HPC tier provides additional,
opportunistic simulation capacity under batch scheduling. These
tiers are connected through a fault-resilient distributed log that
serves as the primary mechanism for data exchange, model
dissemination, and system coordination.

At a high level, \NAME operates as a continuous inference loop.
Edge nodes generate and consume sensor data while performing
inference using the most recent available model. Simulation and
training pipelines consume this data to produce updated models,
which are published asynchronously and deployed at the edge as
they become available. This design enables continuous operation
while hiding HPC delay and reducing model staleness over time.

\subsection{Edge Inference Tier}

The edge tier performs data ingestion, inference, and interaction
with the physical environment. Sensor devices generate streaming
measurements that are transmitted over local wireless networks and
published to the distributed log. This log decouples data producers
from consumers, enabling both local inference and remote processing.

Edge services process incoming 
data using the most recent surrogate model, while
updated models are retrieved and deployed asynchronously. This
allows inference to proceed continuously while incorporating
improved models without interruption.

The edge tier supports heterogeneous networking environments,
including private 5G deployments. Advanced capabilities such as
network slicing enable isolation of inference-critical traffic and
support experimentation with new communication patterns. The
architecture also supports pluggable surrogate models, enabling
evaluation of different modeling approaches under varying resource
constraints.

\subsection{Simulation and Training Pipeline}

The simulation and training pipeline generates updated surrogate
models from sensor data and high-fidelity simulations. Sensor data
from the log is used to parameterize simulations, whose outputs are
used to train surrogate models for low-latency inference.

The pipeline supports multiple simulation frameworks and surrogate
model types, including physics-informed and operator-learning
approaches. Components are integrated through well-defined
interfaces, enabling new models and workflows to be incorporated
without modifying the overall system.

Simulation and training execute on both the dedicated cluster and
HPC tiers. The outputs are versioned models that are published to
the log and deployed asynchronously at the edge.

\subsection{Reverse Backfill on HPC Systems}

\NAME incorporates HPC systems as an opportunistic source of
simulation and model improvement. Rather than requiring predictable
execution, simulation and training jobs are submitted to shared HPC
systems and execute whenever resources become available.

This reinterprets backfilling as a mechanism for improving model
accuracy rather than utilization. Completed jobs produce updated
models that are published and deployed asynchronously, increasing
the frequency of model updates despite irregular execution.

Inference proceeds independently of HPC execution, using the most
recent model available. This decoupling enables the system to
tolerate queue delays and long-running simulations while
maintaining continuous operation.

\subsection{Data and Model Orchestration}

\NAME uses a distributed, fault-resilient log to coordinate data
movement and model deployment across tiers. Sensor data, simulation
inputs, and model artifacts are published to the log, enabling
asynchronous communication among system components.

Simulation pipelines consume data from the log, while updated models
are versioned and published for deployment at the edge. This
mechanism supports model lifecycle management, including versioning,
replacement, and rollback, and ensures that the most recent model is
always available.
By unifying data and model flow within a single abstraction, the log
enables loose coupling, fault tolerance, and flexible integration of
new components across the system.

\section{System Instantiation}
\label{sec:impl}

We instantiate \NAME using a real-world digital agriculture deployment
that couples edge sensing with CFD
simulations to infer airflow within a large agricultural screenhouse.
This deployment serves as a representative realization of the
software architecture described above and enables evaluation under
realistic sensing, networking, and HPC constraints.

\begin{figure}[t]
    \centering
    \includegraphics[width=\linewidth]{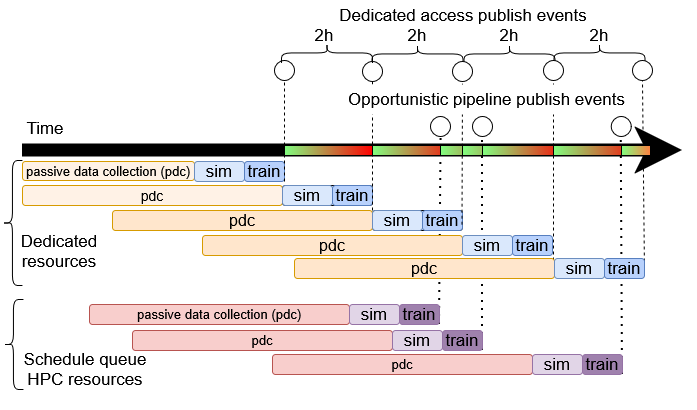}
    \caption{Timeline of the \NAME instantiation illustrating asynchronous,
    simulation-driven model updates. Passive data collection (pdc),
    simulation (sim), and training (train) stages overlap across multiple
    pipeline instances, while model updates are published opportunistically
    upon completion. This design enables continuous inference despite
    irregular and delayed HPC execution.
    \label{fig:pipeline-diagram}
    }
\end{figure}

As shown in Figure~\ref{fig:pipeline-diagram}, the test deployment
implements an asynchronous, simulation-driven pipeline in which data
collection, simulation, and training stages overlap across multiple
pipeline instances, and model updates are published opportunistically
as they complete.  
The goal is to maintain continuous, low-latency inference while
mitigating model staleness under irregular and delayed HPC-driven
model updates.

A set of simulations (\textit{sim}) are launched in parallel on
dedicated resources using data collected up to the time of
execution (denoted \textit{pdc} in Figure~\ref{fig:pipeline-diagram}).
Training begins only after all simulations complete and simulation outputs are 
transformed for training and transferred to training resources (e.g. GPUs).  
Once training finishes,
a new pipeline instance is initiated using the most recent data,
resulting in overlapping pipeline executions.

In our deployment, the combined
\textit{sim} and \textit{train} stages require approximately two hours
on dedicated CPU and GPU resources. Each completed training stage
produces a new surrogate model that is published to the edge. As a
result, models deployed using only dedicated resources may be up to
two hours old.

To reduce model staleness, the pipeline is also executed
opportunistically on shared HPC systems. Jobs are submitted to batch
queues using workflow management system (WMS) and 
parameterized with the most recent data at the
time of execution. Models produced by these runs arrive at the edge
between those generated by the dedicated pipeline, effectively
increasing the rate of model updates despite irregular execution.
This approach increases the frequency of model updates without
requiring predictable execution, directly mitigating the effects of
model staleness.
Because pipeline execution time might be different for WMS and dedicated resources,
an opportunistic model may occasionally arrive after a dedicated model
yet carry an older training cutoff date---the latest sensor timestamp
included in the training data---.
To handle this, the training cutoff date is recorded alongside each model artifact.
Before updating deployed model, the edge system component compares model cutoff date
against that of the currently deployed model and skips update if the
incoming model's cutoff is not strictly newer.
This ensures that the deployed model's training data is monotonically
non-decreasing in freshness, regardless of the order in which jobs from
different resource tiers complete.

\subsection{\NAME Evaluation Facility}

We implement this pipeline using an Evaluation Facility
and application for making continuous 
meteorological inferences for a 4-acre 
Citrus Under Protective Screens (CUPS) facility.  
CUPS is a large ($200 \times 100 \times 6$ meter)
agricultural screenhouse that is designed to protect
citrus trees from the insect that vectors the causal agent of the deadly 
Huanglongbing (HLB) disease (also known as citrus greening).
The screen mesh is fine enough to protect the crop but 
also modifies the environmental conditions under which citrus is grown.  
The application monitors and predicts these conditions to inform, actuate, and control
irrigation planning, spray management, and frost prevention for CUPS deployments.

The deployment consists of a network of sensors that collect
meteorological data, including wind speed, wind direction,
temperature, and humidity outside the CUPS. These measurements are used to
parameterize high-fidelity computational fluid dynamics (CFD)
simulations that model airflow within the screenhouse. The resulting
simulation outputs are used to train surrogate models that enable
low-latency inference at the edge (i.e. on farm).

We use OpenFOAM~\cite{openfoam} to perform CFD simulations, configured to model the
screenhouse geometry and boundary conditions derived from the sensor
data. The prototype 
implements the \textit{sim} stage as 72 parallel OpenFOAM
simulations each configured with the SnappyHexMesh mesh
generator~\cite{_2026_openfoam} and 
PorousSimpleFOAM~\cite{a2026_openfoam} solver, 
to capture the CUPS screen-filtered atmospheric effects
of the CUPS growing conditions across approximately 7 million 
irregularly shaped cells at the centimeter level.

Simulation outputs are used to train multiple surrogate models,
including a physics-informed neural network (PINN)~\cite{pinn,pinn-survey}, 
a Fourier neural operator (FNO)~\cite{fno}, 
and a principal component regression (PCR)~\cite{pcr} model.
This instantiation leverages the pluggable design of \NAMENS, enabling
multiple surrogate models to be trained and evaluated under identical
system conditions. The models vary in complexity and computational
cost, enabling evaluation of accuracy–latency tradeoffs at the edge.

Trained models are deployed to edge devices 
(Raspberry Pi single board computers~\cite{raspi}), 
where they provide
continuous predictions of airflow across the screenhouse. Sensor
data is streamed to the distributed log, where it is consumed by
both the simulation pipeline and edge inference services. As new
models are produced, they are published to the log,
allowing the system to update predictions without
interrupting operation.

\subsection{The \NAME Data Mover}

\NAME currently assumes that data is transferred between computational 
components (including components executing at the edge) using files.  
For example, the output data from the simulation stage will be a set 
of files that (possibly after a transformation) will be used as 
input files to the training stage.

To enable this, we extended the open-source, CSPOT log-based
event system~\cite{cspot19}  with 
a basic file transfer protocol that 
implements delay-tolerant communication of files between
components. The \NAME Data Mover (RBFDM) uses CSPOT logs to create versioned
files that are created by a file ``push'' to a log and read using a file
``pull.''  By placing the data repositories where the CSPOT logs are stored
outside of all firewalls, any file producer pushes the file that has been
created to a CSPOT log in a repository where it can be fetched by the file's
consumers.

CSPOT assigns a unique sequence number to each log entry.  The RBFDM writes
blocks into a CSPOT log, one at a time, and records the sequence number
associated with the starting and ending log entries and associates these with
a file version number.  A reader of the file can specify the specific version
number and/or the starting sequence number in the log.  The RBFDM API includes
a call that returns the most recent (latest) file version.  To implement a
notification that a new file has been appended to the log, readers poll  the
log looking for an updated file version.

\begin{figure*}[t]
    \centering
    \includegraphics[width=0.9\linewidth]{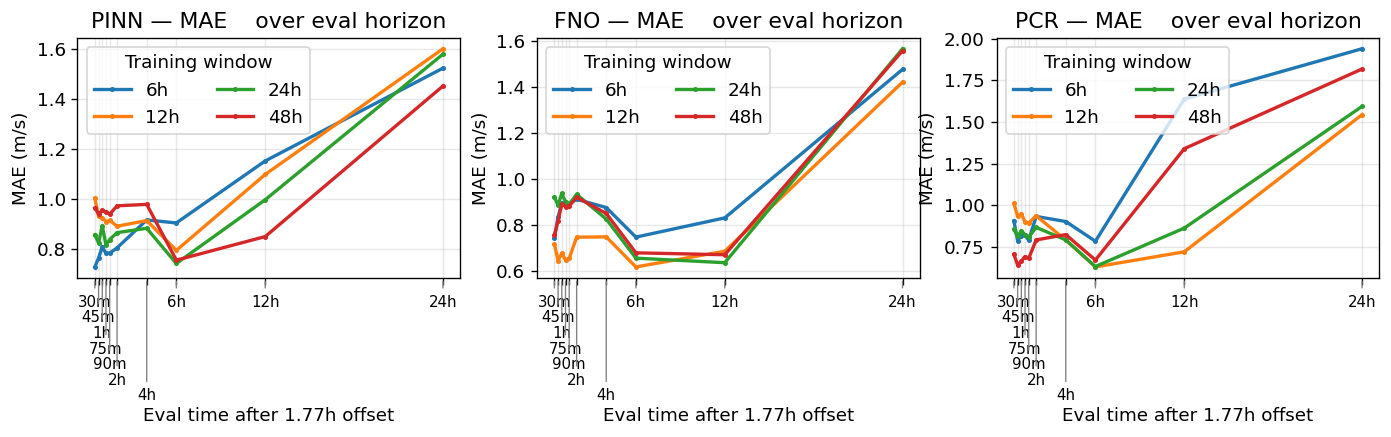}
    \caption{Model accuracy decay over time using different
    history windows for all three models (PINN, FNO, PCR). The x-axis
    shows elapsed time ranging from 30 minutes to 4 hours after model
    generation. The y-axis shows cumulative MAE (m/s) computed by
    aggregating all prediction errors across the three sensor test locations in the field. Each
    subplot shows the decay curves for one model type.}
    \label{fig:accuracy_comparison_3models}
\end{figure*}

All \NAME components that use files as inputs or generate files as outputs
(including edge devices implementing fast inference models) use the RBFDM
facility to read and store their files, respectively.  Further, the \NAME
software stack uses the RBFDM to implement software updates.  When a software
change must be propagated across a deployment, the software (which is
statically linked) is pushed to CSPOT logs by the CI/CD system and then pulled
to each node and installed when the version number updates. The RBFDM uses
CSPOT CAPlets~\cite{caplets21} to implement secure communication.

\subsection{Private 5G Network}
The \NAME private 5G network is built on two open source components: srsRAN
Project~\cite{srsran} for the Radio Access Network (RAN) and Open5GS~\cite{open5gs}
for the 5G core. Both are deployed as Docker containers on a dedicated host machine,
communicating over an internal bridge network.

The gNB is driven by an Ettus Research USRP B210 software defined radio operating on
3GPP Band~n78 with a 40~MHz channel bandwidth and 30~kHz subcarrier spacing in TDD
mode. The core network implements a full 5G Standalone (SA) architecture, including the
AMF, SMF, UPF, NSSF, and supporting network functions. Together, these provide a
self-contained cellular network with no dependency on external carrier infrastructure.

Two Raspberry Pi~4 Model~B single board computers serve as user equipment (UEs), each
fitted with a Quectel 5G modem connected over USB. The modem exposes a QMI
control interface and AT command ports to the host operating system. Each modem carries
a programmable SIM provisioned with a unique IMSI and authentication credentials
registered in the Open5GS subscriber database. On power up, the modem attaches to the
NR~SA cell and establishes a PDU session, which surfaces as a \texttt{wwan0} network
interface on the Raspberry Pi with a statically assigned IP address.

The two Raspberry Pi nodes serve distinct roles in the pipeline, each defining a
separate data path across the RAN. The \emph{sensor data path} carries telemetry
from the sensor node to the inference pipeline, where consistent low latency is
critical for real time operation. The \emph{model distribution path} delivers
surrogate model weight updates from the cloud backend to the edge inference node,
a flow that demands high throughput but tolerates moderate delay. Since these
two paths share the same radio link, contention between them can degrade either latency or throughput. This motivates the use of network slicing to isolate the two flows.

This instantiation demonstrates how \NAME supports
simulation-driven applications in which model updates are delayed
and irregular, while enabling continuous inference through
asynchronous and opportunistic model improvement.

\section{Empirical Evaluation}
\label{sec:eval}
We next use the \NAME Evaluation Facility and CUPS application 
to empirically evaluate the system end-to-end. We demonstrate the effectiveness of \NAME by presenting
our results in three stages. First, we establish the operational cadence
of our dedicated-access computational pipeline. Second, we show how model
accuracy decays over time, detailing the effects of ``backfilling'' the
production of inference models by the dedicated pipeline with opportunistic
computations run at different sites.
Finally, we evaluate the deployment of the inference models onto
edge devices, showing that they can be delivered and
executed with minimal overhead.

\subsection{Dedicated-Access Pipeline Performance}
\label{sec:dedicated}
As described in the previous section, \NAME
is designed to produce an inference model on a regular, maximum cadence
using a limited, but dedicated cluster and set of two GPUs (which we refer to as
the ``dedicated-access pipeline'').
It then augments this maximal model-production cadence
with additional computations that are initiated after 
waiting (at different sites) in a batch queue
for an unpredictable 
time period.  
As long as the models produced by dedicated resources remain
sufficiently accurate to support inference, opportunistic model
updates serve only to improve prediction accuracy.

\begin{figure*}[ht]
    \centering
    \includegraphics[width=0.9\linewidth]{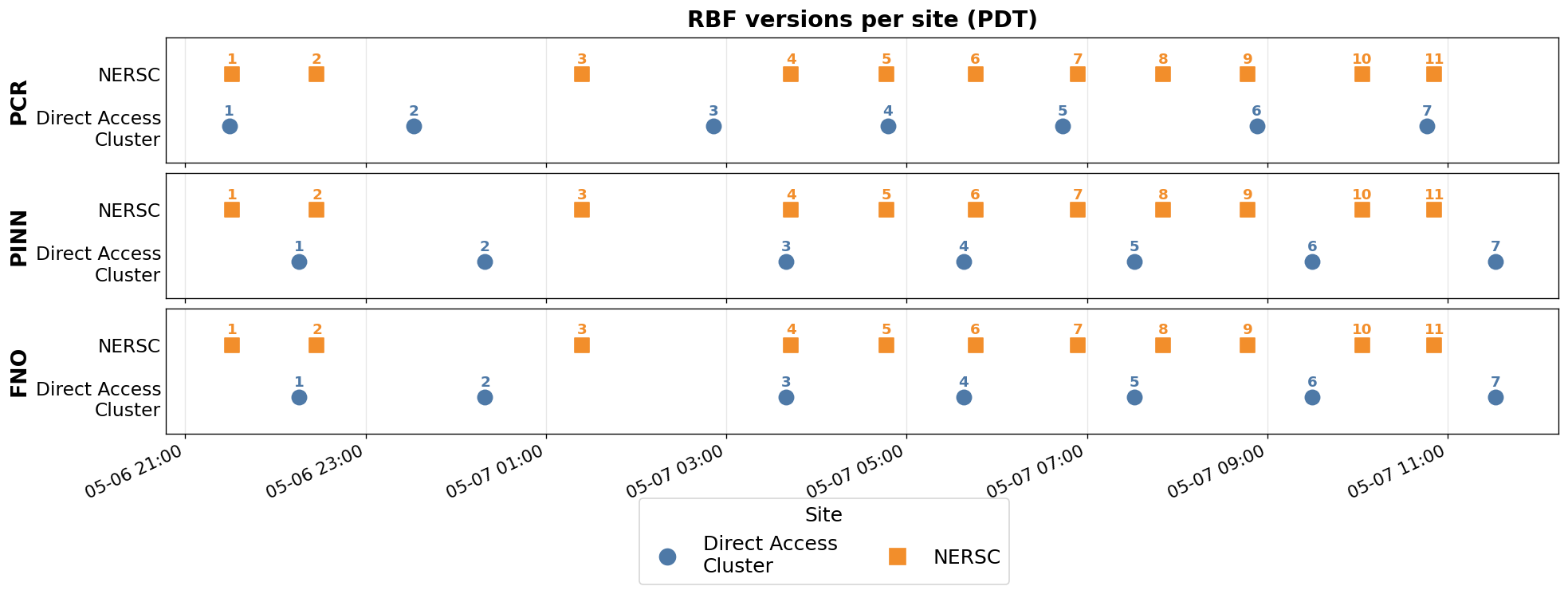}
    \caption{Timeline of model publish events for all three model types (PINN, FNO, PCR) during a simultaneous live experiment on two resources: the dedicated-access cluster and the NERSC batch-queue HPC.}
    \label{fig:combined-cadence}
\end{figure*}

On average across 15 runs, the dedicated-access pipeline 
completes in $134.8 \pm 58.0$ minutes. Of this total, the \textit{sim} stage accounts
for approximately 52 minutes for CFD computation (72 nodes) 
and 14 minutes for data transformation of simulation outputs (for use
in model training).  
The \textit{train} stage executes PINN, FNO, and PCR training in
parallel on CPU/GPU resources and requires approximately 55 minutes in
total. Individually, PINN training requires $50.0 \pm 21.6$ minutes,
FNO requires $54.8 \pm 18.2$ minutes, and PCR requires
$15.9 \pm 3.4$ minutes.
The remaining time 
is due to data fetching from the edge, model transfer to the edge, logging, 
and other miscellaneous system overhead.
Thus, a new inference model is available at the edge, on average, every
134.8 minutes. 
This end-to-end cadence represents the fastest achievable
retraining cycle with dedicated resources, completely independent of HPC
queue contention.

The $\pm 58.0$ minute standard deviation (42\% of the mean) arises from two
sources: (i) synchronization overheads during data staging and model transfer
across the cluster, and (ii) variable training iteration counts when the 72
simulation cases produce different numbers of unique data points. Despite this
inherent variability, the dedicated-access pipeline remains operationally
predictable and quantifiable, enabling reliable operational planning.

\subsection{Inference Model Accuracy Decay}
\label{sec:decay}

Because the inference models are trained from simulations using ``live''
data and data immediately preceding it from the recent past, they are most 
accurate for the immediate
future and their accuracy decays with time.  In this study, the dedicated pipeline produces 
an inference model that is trained from data that is (on average) 
at least 134.8 minutes old -- the data that was available when the simulations were initiated.

Given this 134.8-minute dedicated pipeline cadence, the critical questions are: 
\begin{itemize}
\item What accuracy do the deployed inference models maintain when
produced at the maximal cadence?
\item Does retraining more frequently improve accuracy and, if so, by how much?
\end{itemize}
Figure~\ref{fig:accuracy_comparison_3models} displays the decay of model
accuracy over time for all three models (PINN, FNO, PCR). 
In the figure, the x-axis shows the time, in hours,
since the model has been produced, and the y-axis shows the mean absolute error (MAE)
in meters per second when a prediction generated by a model is compared to
a measurement at a test point in the field.  That is, to compute MAE, we parameterize
the inference model with the current data and compare the inference it makes to
the next available measurement.  Thus the figure depicts the ``decay'' in accuracy (measured as MAE)
as the time since the model was produced with respect to different test
points in the field.

Note that the CFD simulations all take the latest real-time sensor data and
a short ``history'' of measurements immediately preceding it when they are initiated.
In Figure~\ref{fig:accuracy_comparison_3models} we show the accuracy decay for each
model using a different color, depending on history length.  These results show the
opportunity for dynamic and automatic hyper-parameter tuning.  

For example, PINN is
most accurate between zero and 6 hours after the model was generated when the simulations
were trained with either 6 hours (blue) or 24 hours (green) of history
length. Note that the accuracy curves for PINN cross at the 6 hour mark and
afterwards, 48-hours of history (red) is most accurate.  Thus, if an opportunistic
computation can produce an additional model in under 6 hours, it should use
a 6-hour history length.  If not, it should use a 48-hour history length.  However,
the length of history affects the training time (but not the simulation time) and this increases the
maximal pipeline cadence.  Finding
the optimal history length (as a hyper-parameter) dynamically,
for each computation is the subject of our future work.  In this study, we use a 6-hour
history length (the blue features in the figure) for all simulations which is minimal for PINN and PCR but which
is approximately 0.1 meters per second less accurate (in terms of MAE) compared to a 12-hour
history for FNO.

\subsection{Opportunistic Model Regeneration}

Each new model generation effectively resets the system to time zero on the
decay curve —- the point of maximum accuracy. Using dedicated resources, fresh
models are available every 134.8 minutes (the \textit{sim}+\textit{train} time in Figure~\ref{fig:pipeline-diagram}).
Each additional
generation, regardless of source, shifts the operational system's current point
leftward on the decay curve, immediately improving prediction quality.

Note that without the ability to predict batch queue delays, model generation
events occur at times that are randomly distributed between the maximal
cadence generations.  On average, then, if jobs submitted to other HPC sites
generate one additional inference model within each
of the maximal cadence periods, the decay period is cut in half to approximately 67 minutes.
Two generations within a maximal cadence period cuts
the decay period to $1/3$ of the maximal period (45 minutes), three
generations cut it to $1/4$ (34 minutes) and so on.

In the test setting, new data is available every 5 minutes and the sensor
measurement error
for wind speed at the test points is between $\pm 0.44$ and $\pm 0.87$ 
meters per second.  These physical parameters set an upper bound on the effectiveness of \NAME.  There is no possible improvement in accuracy decay when a model
is generated faster than every 5 minutes (which is 20 additional model generations per
a maximal cadence period of 134.8 minutes) and an inference error below $0.44$ meters per second is 
indistinguishable from measurement error.

We validate this analysis with a live multi-site experiment.
Figure~\ref{fig:combined-cadence} shows model publish events for PCR, PINN, and FNO on a shared
time axis spanning the night of May~6--7, 2026. Blue circles denote dedicated cluster publish events and orange
squares denote NERSC publish events. Dedicated cluster events appear at regular approximately 2-hour intervals
across all model types; PCR events are offset from PINN and FNO because PCR trains on CPU and
converges more quickly. NERSC events are less frequent: each batch job allocation runs until its time limit expires, after
which a new job must be submitted to the queue. The average wait for a new allocation is
approximately 17 hours, creating mandatory gaps of at least 18 hours during which no NERSC publish
events are possible. For these experiments, we observed queue wait times for the NERSC Perlmutter 
HPC system to be 17-19 hours for the 72-CPU jobs and 11-38 minutes for the 2-GPU jobs~\cite{nersc-perlmutter-waittimes}.

The patterns displayed in the figure demonstrate the 
complementary nature of the resource types: dedicated cluster provides a
steady, predictable cadence while NERSC contributes opportunistic events that fall between dedicated cluster
generations. Together, they reduce the average time between consecutive publish events by nearly
half, keeping effective model age below 2 hours on the decay curve of
Figure~\ref{fig:accuracy_comparison_3models} and placing all three model types below the
0.88~m/s sensor measurement error bound.
The longer cadence events for both resources are due to wind condition differences, which lead to 
bigger variability in wind speed and consequently longer training windows.

Table~\ref{tab:combined-cadence} reports publish-event latency statistics for FNO models across all
three resource configurations. The dedicated cluster alone yields an average interval of 134.8 minutes, consistent
with the $\sim$2-hour dedicated pipeline cadence. NERSC alone achieves a shorter average of 80.0 minutes
per completed job, but with high variability (std 40.4 min) from unpredictable queue waits. The
combined configuration further reduces both the average (50.0 min) and the minimum (3.3 min)
inter-publish interval.
The 2.7$\times$ reduction in average cadence (134.8 to 50.0 minutes) confirms that opportunistic
HPC contributions materially reduce model staleness even under irregular availability.
As discussed in Section~\ref{sec:impl}, a dedicated-pipeline job that
began computation earlier can sometimes finish after an opportunistic NERSC job
that started later, because NERSC per-job execution times are shorter.
In these cases, the edge device compares the training cutoff date of
the incoming model against that of the currently deployed model and
skips deployment if the incoming cutoff is not strictly newer.
This check introduces negligible overhead, and in the current
experimental setup the execution-time discrepancy between the dedicated
pipeline and NERSC is small enough that such ordering inversions are
rare.
Overall, these results demonstrate that \NAME's design effectively leverages both dedicated and opportunistic resources
to maintain a high cadence of model updates, directly mitigating the effects of model staleness and improving inference accuracy.

\begin{table}[t]
    \centering
    \caption{Minutes between FNO model publish events}
    \label{tab:combined-cadence}
    \vspace{0.05in}
    \footnotesize
    \begin{tabular}{l|l|l|l|l}
        \hline
        \textbf{Combination} & \textbf{Min}& \textbf{Avg}& \textbf{Max} & \textbf{Std}\\
        \hline                                     
            dedicated cluster          &      113.4  &     134.8   &    200.4  &  32.9\\
            NERSC            &      47.9    &     80.0   &    176.5   &  40.4\\
            dedicated cluster + NERSC     &     3.3   &     50.0    &   135.8  &  34.3\
    \end{tabular}
\end{table}

\subsection{Edge Performance Summary}

\begin{figure}[h]
    \centering
    \includegraphics[width=\columnwidth]{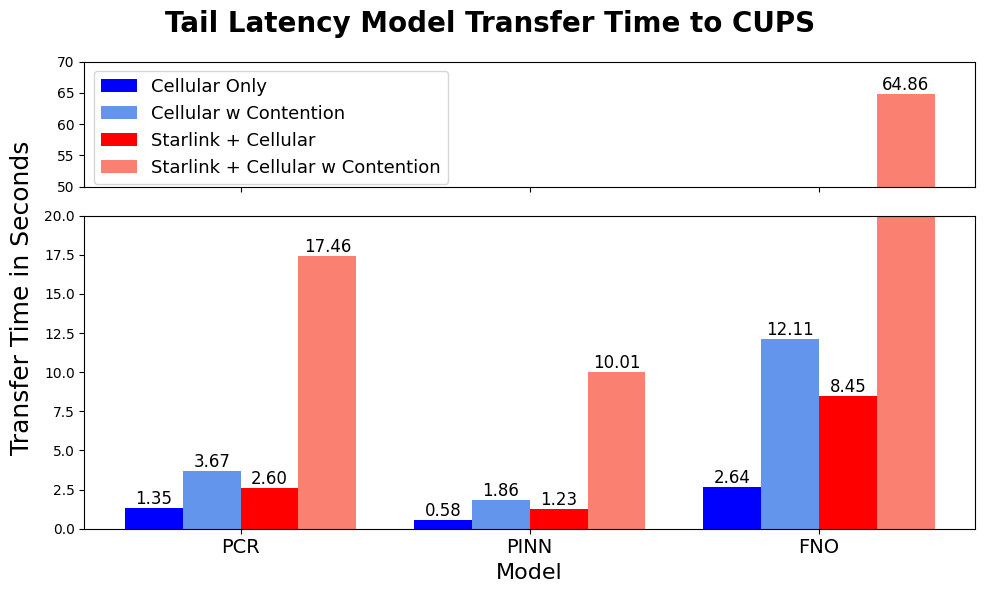}
    \caption{P-95 Model Transfer Time of 100 runs. Transfer time is in seconds. Each bar represents a different network path (cell only or starlink and cellular) for model delivery. }
    \label{fig:cups_result}
\end{figure}

We next evaluate the impact of model transfer latency and edge inference
performance to determine whether these components affect end-to-end
system behavior. 
Figure~\ref{fig:cups_result} compares model transfer times (P-95 values for worst-case tail latency)
for different network conditions on the CUPS deployment. 
Overall, model transfer times correspond to model size: the PINN model is 290KB,
the FNO model is 9.1MB, and the PCR model is 1.1MB.

In Table~\ref{tab:throughput_slicing}, we summarize the throughput degradation per model
with and without network slicing in an indoor 5G private network deployment supporting
customized network configurations.
FNO shows the starkest difference due to its size: Without slicing, 
throughput drops from 4.92 to 3.88\,MB/s (21\% degradation), 
while with slicing it holds at 4.62\,MB/s (2\% change).

\begin{table}[h]
    \centering
    \caption{Model download throughput (MB/s) with and without network slicing.
    Values are means over 100 runs. Degradation is the percentage change
    under contention.}
    \label{tab:throughput_slicing}
    \vspace{0.05in}
    \footnotesize
    \begin{tabular}{l c c c c c c}
        \hline
        & \multicolumn{3}{c}{\textbf{No Network Slicing}} & \multicolumn{3}{c}{\textbf{Network Slicing}} \\
        \cmidrule(lr){2-4} \cmidrule(lr){5-7}
        \textbf{Model} & \textbf{Iso.} & \textbf{Cont.} & \textbf{Deg.} & \textbf{Iso.} & \textbf{Cont.} & \textbf{Deg.} \\
        \hline
        PCR  & 2.68 & 2.15 & $-$20\% & 2.67 & 2.50 & $-$6\% \\
        PINN & 1.37 & 1.06 & $-$23\% & 1.28 & 1.31 & $+$2\% \\
        FNO  & 4.92 & 3.88 & $-$21\% & 4.72 & 4.62 & $-$2\% \\
        \hline
    \end{tabular}
\end{table}

Across all configurations, model transfer times are
small relative to the simulation–training pipeline latency (on the
order of hours). Even under constrained network conditions, transfer
latency remains low, indicating
that communication does not constitute a bottleneck in the system.
Finally, our results indicate that 5G network slicing shields model download throughput effectively under contention.

We evaluate inference latency on resource-constrained edge
devices (omitted here due to space constraints). 
We find that all surrogate models execute within a few seconds, with
lightweight models (e.g., PCR) achieving sub-second latency. More
complex models (e.g., PINN, FNO) remain well within practical
operational bounds. These results demonstrate that edge inference is
feasible across a range of model types and introduces negligible
overhead relative to model update times.

Overall, these results indicate that system performance is dominated
by simulation and training latency, and that networking and edge
inference do not limit end-to-end performance. This reinforces the
design focus of \NAME on asynchronous, simulation-driven model
improvement, where reducing model staleness through opportunistic
HPC execution provides the primary gains. The results also demonstrate
that \NAME facilitates the integration and evaluation of diverse
surrogate models and networking configurations, enabling flexible
experimentation across model types and communication environments
without impacting system operation.

\section{Related Work}
\label{sec:related}
Early versions of the \NAME concept, including the online--offline
learning and inference loop, were explored in a prior workshop
paper~\cite{xgfabric-xloop}. The work described herein 
substantially extends that effort by
formalizing the architecture, introducing reverse backfill for
asynchronous HPC integration, and providing a full system
implementation and evaluation.

Recent work has explored integrating distributed execution and
inference pipelines across heterogeneous resources. Systems such as
Ray~\cite{ray} and InferLine~\cite{inferline} provide abstractions for
distributed training and serving, while edge–cloud platforms such as
Jellyfish~\cite{jellyfish} and EC5~\cite{ec5} optimize inference
placement and resource utilization under latency and resource
constraints. Additional work on distributed inference has explored
model partitioning and resource-aware adaptation~\cite{neurosurgeon,edgent,jointdnn},
as well as serverless execution across the computing continuum~\cite{loconte2024serverless}.
However, these systems assume pre-trained models and tightly coupled
execution, and do not incorporate delayed or asynchronous model updates
from remote HPC resources. In contrast, \NAME integrates
simulation-driven training into the inference pipeline, treating HPC as
an asynchronous model-improvement engine within a continuous inference
loop.

Continuous learning systems assume that model updates can be derived
directly from streaming observations~\cite{wang2024continual,gomes2019streaming}. 
\NAME targets simulation-driven settings in which updates depend on
expensive, batch-scheduled computation, requiring explicit handling of
delayed and irregular model refresh.

Simulation-driven learning approaches use high-fidelity simulations to
generate training data for surrogate models, including physics-informed
neural networks and operator-learning methods such as FNOs and
DeepONets~\cite{pinn,fno,deeponet}, with applications in computational
fluid dynamics and related domains~\cite{cfd-for-dl-training,pinn-and-cfd}.
Recent work has also explored scientific machine learning and
HPC-integrated training workflows~\cite{lbann,exalearn}. While these
approaches improve model accuracy, they focus primarily on model design
and training rather than deployment across distributed systems. \NAME
complements this work by integrating simulation, training, and
deployment into a unified pipeline spanning edge, cloud, and HPC
resources, enabling continuous retraining and deployment under
variable system conditions.

Distributed workflow systems provide mechanisms for orchestrating
large-scale computation across clusters, clouds, and HPC systems,
including pilot-job systems and scientific workflow
engines~\cite{laminar,pegasus,radicalpilot,swift-t}. These systems
primarily target batch-oriented workloads and do not support continuous,
closed-loop inference with delayed model updates. Prior work has also
explored integrating ML pipelines with distributed orchestration, but
remains largely batch-oriented~\cite{kubeflow,tfx}. At the systems
level, recent efforts have demonstrated asynchronous in-situ surrogate
training on HPC~\cite{balin2023insitu} and urgent HPC simulation driven
by live data~\cite{flatken2023vestec}, providing key building blocks
for HPC–AI integration. \NAME builds on these ideas by enabling
asynchronous, simulation-driven model evolution within a continuous
edge inference loop.

ROSE~\cite{roseRADICAL} explores orchestration for surrogate training
using active learning, while \NAME focuses on maintaining continuous
inference under delayed model updates. Finally, \NAME leverages
private 5G connectivity to support reliable model distribution to edge
devices~\cite{xgfabric-xloop,polese2024openran,sarah2023mec5g,geng2024private5g,zhang2024tsn,satka2023tsn5g}.

In summary, prior work has explored edge inference, surrogate modeling,
and distributed workflows largely in isolation. \NAME differs by
treating HPC as an asynchronous model-improvement engine within a
continuous inference loop, enabling low-latency prediction despite
delayed model updates.

\section{Conclusions and Future Work}
\label{sec:conc}
We presented \NAMENS, a hybrid edge–HPC architecture for simulation-driven
systems in which model updates are bounded by high-fidelity computation.
By decoupling inference from simulation and training, \NAME enables
continuous, low-latency inference while incorporating improved models
asynchronously as they become available.

Our results show that system performance is dominated by simulation
latency, with networking and edge inference contributing negligible
overhead. This demonstrates that HPC resources can be effectively
integrated into operational systems as asynchronous model-improvement
engines, enabling practical deployment of simulation-driven learning, with a
2.7$\times$ reduction in average model staleness for the deployments studied.
\NAME also provides a flexible platform for evaluating diverse models
and networking configurations across heterogeneous environments.
This research is supported in part by NSF awards CNS-2444318 and CNS-2107101,
and DoE award DE-SC0025541

\bibliographystyle{IEEEtran}
\bibliography{common}

\end{document}